\documentclass{French}
\usepackage{latexsym}
\usepackage[french]{babel}
\usepackage{graphicx}
\usepackage{subfigure}

\begin{document}
\subh{Discussing an intriguing result on model-free control}
\author{C\'{e}dric Join\ts{1,5,6}, Emmanuel Delaleau\ts{2}, Michel Fliess\ts{3,5}, Claude  H. Moog\ts{4}}
\aff{\removelastskip 
\ts{1}CRAN (CNRS, UMR 7039), Université de Lorraine, BP 239, 54506 Vand{\oe}uvre-lès-Nancy, France. {\tt Cedric.Join@univ-lorraine.fr}\\
\ts{2}Département de Mécatronique, \'Ecole nationale d'ing\'{e}nieurs de Brest, 29280 Plouzan\'{e}, France. {\tt emmanuel.delaleau@enib.fr} \\
\ts{3}LIX (CNRS, UMR 7161), \'Ecole polytechnique, 91128 Palaiseau, France. {\tt Michel.Fliess@polytechnique.edu}\\
\ts{4}LS2N (CNRS, UMR 6004), 44321 Nantes 03, France.  {\tt moog@ieee.org} \\
\ts{5}AL.I.E.N. (ALgèbre pour Identification \& Estimation Numériques), 7 rue Maurice Barrès, 54330 Vézelise, France. \\ {\tt \{michel.fliess, cedric.join\}@alien-sas.com}\\
\ts{6}Projet Non-A, INRIA Lille -- Nord-Europe, France}

\resu{Un exemple mathématique élémentaire prouve, gr\^{a}ce au critère de Routh-Hurwitz, un résultat à l'encontre de la pratique actuelle en commande sans modèle: il peut y avoir plus de difficultés à régler un correcteur proportionnel \og{intelligent}\fg \ (iP) qu'un proportionnel-dérivé intelligent (iPD). Les simulations numériques de l'iPD et d'un PID classique tournent largement en faveur du premier. Introduction et conclusion analysent la commande sans modèle à la lumière des avancées actuelles.}
\abstract{ An elementary mathematical example proves, thanks to the Routh-Hurwitz criterion, a result that is intriguing with respect to today's practical understanding of model-free control, \textit{i.e.}, an ``intelligent'' proportional controller (iP) may turn to be more difficult to tune than an intelligent proportional-derivative one (iPD). The vast superiority of iPDs  when compared to classic PIDs is shown via computer simulations. The introduction as well as the conclusion analyse model-free control in the light of recent advances.}
\moi{\textbf{\textcolor{AbsBlue}{MOTS-CL\'{E}S.}}\hphantom{--} Commande sans modèle, correcteurs iP, correcteurs iPD, PID, critère de Routh-Hurwitz, réparation, apprentissage, intelligence artificielle.}
\kwd{Model-free control, iP controllers, iPD controllers, PID, Routh-Hurwitz criterion, fault accommodation, machine learning, artificial intelligence.
}

\chapter{Un résultat intrigant en commande sans modèle}

\begin{flushright}
La première fois qu'Aurélien vit Bérénice, il la trouva franchement laide. \\
Aragon (\textit{Aurélien}. Paris: Gallimard, 1944)
\end{flushright}

\section{Introduction}
\subsection{Généralités}
Les faits suivants sont connus de tout automaticien:
\begin{itemize}
\item \'Ecrire un \og{bon}\fg \ modèle mathématique d'une machine réelle, c'est-à-dire non idéalisée, comme en physique fondamentale, est redoutable, voire impossible. Ainsi s'explique la popularité industrielle stupéfiante des correcteurs PID (voir, par exemple, \cite{astrom}, \\ \cite{franklin}, \cite{janert}, \cite{lunze}, \cite{od}, \\ \cite{rotella}). Une telle modélisation y est sans objet.
\item Le tribut est lourd:
\begin{itemize}
\item performances médiocres,
\item défaut de robustesse,
\item réglage laborieux des gains.
\end{itemize}
\end{itemize}
La \og{commande sans modèle}\fg, ou {\it model-free control}, \cite{ijc13} et ses correcteurs \og{intelligents}\fg \ ont été inventés pour combler ces lacunes. De nombreuses publications récentes, dans les domaines les plus divers, démontrent leur efficacité et simplicité, non seulement en France, mais aussi, et même davantage, à l'étranger: voir, par exemple, les références de \cite{ijc13}, et  \newline \cite{alinea}, \cite{bldg} et leurs références.

La bibliographie de \cite{ijc13} atteste que la dénomination \emph{model-free control} appara\^{\i}t maintes fois dans la littérature, mais en des sens distincts du nôtre. L'importance croissante de l'intelligence artificielle et de l'apprentissage, au travers des réseaux de neurones notamment, s'est fort naturellement greffée au sans-modèle: voir, par exemple, \cite{kr}, \cite{li}, \cite{luo}, \cite{nature}, \cite{radac0}, \cite{radac}. Nos techniques, sans nul besoin de calculs lourds \cite{nice}, éludent cette tendance actuelle de l'informatique (voir \cite{brest}, \cite{toulon}, \cite{ieee17} pour des illustrations concrètes).  

\subsection{Bref aperçu de la commande sans modèle\protect\footnote{Pour plus de détails, voir \cite{ijc13}.}}
On remplace le modèle global inconnu par le modèle \emph{ultra-local}:
\begin{equation}
y^{(\nu)} = F + \alpha u \label{1}
\end{equation}
\begin{itemize}
\item Les variables $u$ et $y$ désignent respectivement la commande et la sortie.
\item L'ordre de dérivation, choisi par l'ingénieur, $\nu \geq 1$ est $1$, en général. Parfois, $\nu = 2$. On n'a jamais rencontré $\nu \geq 3$ en pratique.
\item L'ingénieur décide du paramètre $\alpha \in \mathbb{R}$  de sorte que les trois termes de \eqref{1} aient même magnitude. Une identification précise de $\alpha$ est, donc, sans objet.
\item On estime $F$ grâce aux mesures de $u$ et $y$.
\item $F$ subsume non seulement la structure inconnue du système mais aussi les perturbations externes\footnote{Cette distinction entre structure interne et perturbations externes se retrouve partout. Elle ne présente, à notre avis, aucune évidence \textit{a priori}. Les confondre est une percée conceptuelle indubitable. Comparer avec \og{la commande par rejet actif de perturbations}\fg, ou \textit{Active Disturbance Rejection} (\textit{ADRC}) (voir, par exemple, \cite{sira}).}.
\end{itemize}
Si $\nu = 2$, on ferme la boucle avec un régulateur \emph{intelligent proportionnel-intégral-dérivé}, ou \emph{iPID}, c'est-à-dire une généralisation des PID classiques,
\begin{equation}\label{ipid}
u = - \frac{F_{\rm estim} - \ddot{y}^\ast - K_P e - K_I \int e - K_D 
\dot{e}}{\alpha}
\end{equation}
\begin{itemize}
\item $F_{\rm estim}$ est une estimée $F$.
\item $y^\ast$ est la trajectoire de référence.
\item $e = y^\ast - y$ est l'erreur de poursuite.
\item $K_P, K_I, K_D \in \mathbb{R}$ sont les gains.
\end{itemize}
Il vient, d'après \eqref{1} et \eqref{ipid},
\begin{equation}\label{track}
\ddot{e} + K_D \dot{e} + K_P e + K_I \int e = F_{\rm estim} - F
\end{equation}
On obtient une \og{bonne}\fg \ poursuite si l'estimée $F_{\rm estim}$ est \og{bonne}\fg, c'est-à-dire $F - F_{\rm estim} \simeq  0$. Contrairement aux PID classiques, \eqref{track} prouve la facilité, ici, du choix des gains.

Si $K_D = 0$ on a un régulateur \emph{intelligent proportionnel-intégral}, ou \emph{iPI},
\begin{equation*}\label{ipi}
u = - \frac{F_{\rm estim} - \ddot{y}^\ast - K_P e - K_I \int e}{\alpha}
\end{equation*}
Si $K_I = 0$ on a un régulateur \emph{intelligent proportionnel-dérivé}, ou \emph{iPD},
\begin{equation}\label{iPD}
u = - \frac{F_{\rm estim} - \ddot{y}^\ast - K_P e - K_D \dot{e}}{\alpha}
\end{equation}
Le plus fréquemment, $\nu = 1$.  On obtient alors un régulateur \emph{intelligent proportionnel}, ou \emph{iP},
\begin{equation}\label{ip}
u = - \frac{F_{\rm estim} - \dot{y}^\ast - K_P e}{\alpha}
\end{equation}

\noindent{\bf Remarque}. Voir \cite{delaleau} pour une autre approche de la stabilisation.

Voici deux exceptions où un iPD est employé avec $\nu = 2$: \cite{comp}, \\ \cite{ieee17}\footnote{La littérature sur les illustrations du sans-modèle contient plusieurs exemples avec emploi d'un iPID et $\nu = 2$, mais sans justification aucune, comme l'absence de frottements \cite{ijc13}. Un iP avec $\nu = 1$ aurait suffi peut-être. D'où une implantation encore plus simple.}. 

\subsection{But}
Cet article exhibe un exemple linéaire, $\ddot{y} - \dot{y} = u$, \textit{a priori} élémentaire, où un iPD doit remplacer un iP, contrairement à ce que l'on aurait pu croire na\"{\i}vement. L'explication repose sur le critère bien connu de Routh-Hurwitz (voir, par exemple, \cite{gant}). Il démontre l'\og{étroitesse}\fg \ de l'ensemble des paramètres stabilisants $\{\alpha, K_P\}$ en \eqref{ip}. 

\subsection{Plan}
Le paragraphe {\ref{example}} présente notre exemple et les excellents résultats obtenus avec un iPD. Au paragraphe suivant, les difficultés rencontrées avec un iP sont expliquées gr\^{a}ce au critère de Routh-Hurwitz. L'équivalence démontrée en \cite{pId}, \cite{ijc13} entre PID et iPD nous conduit à comparer leurs performances au paragraphe {\ref{PD}}: l'avantage des iPD y est manifeste. On en déduit en conclusion quelques pistes de réflexions sur les correcteurs intelligents associés au sans-modèle.

\section{Notre exemple}\label{example}
\subsection{Présentation}\label{pres}
Soit le système linéaire, stationnaire et instable,
\begin{equation}\label{x}
\boxed{\ddot{y}-\dot y=u }
\end{equation}
D'après \eqref{1}, il vient, si $\nu = 1$,
\begin{equation}\label{Fana}
F=-\alpha \ddot y+(1+\alpha)\dot y
\end{equation}
L'iP déduit des calculs de \cite{ijc13} fonctionne mal.

\subsection{iPD}\label{IPD}
\begin{figure*}
\begin{center}
\subfigure[Commande]{
\resizebox*{7.80cm}{!}{\includegraphics{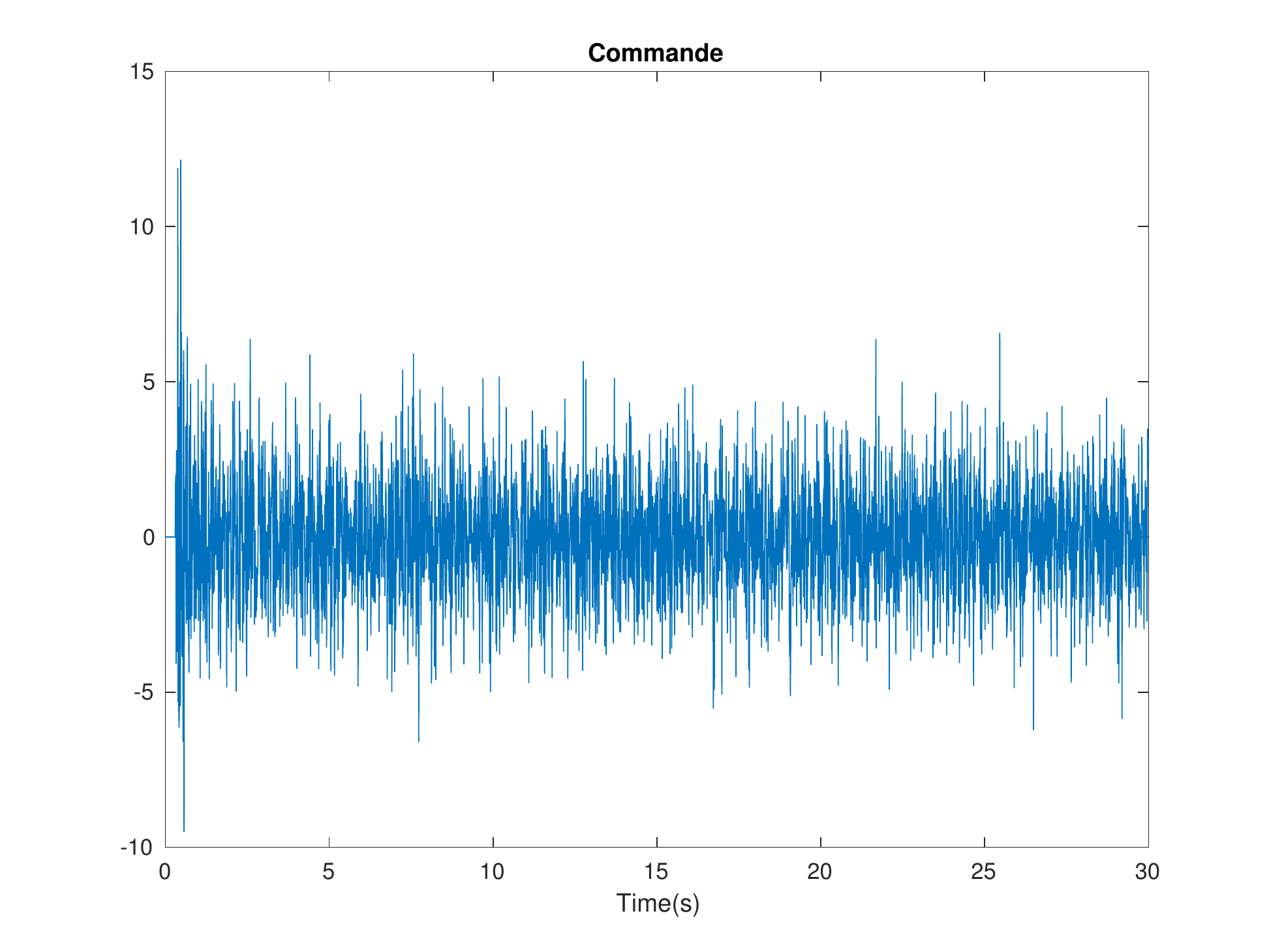}}}%
\subfigure[Sortie, trajectoire de référence (- -)]{
\resizebox*{7.80cm}{!}{\includegraphics{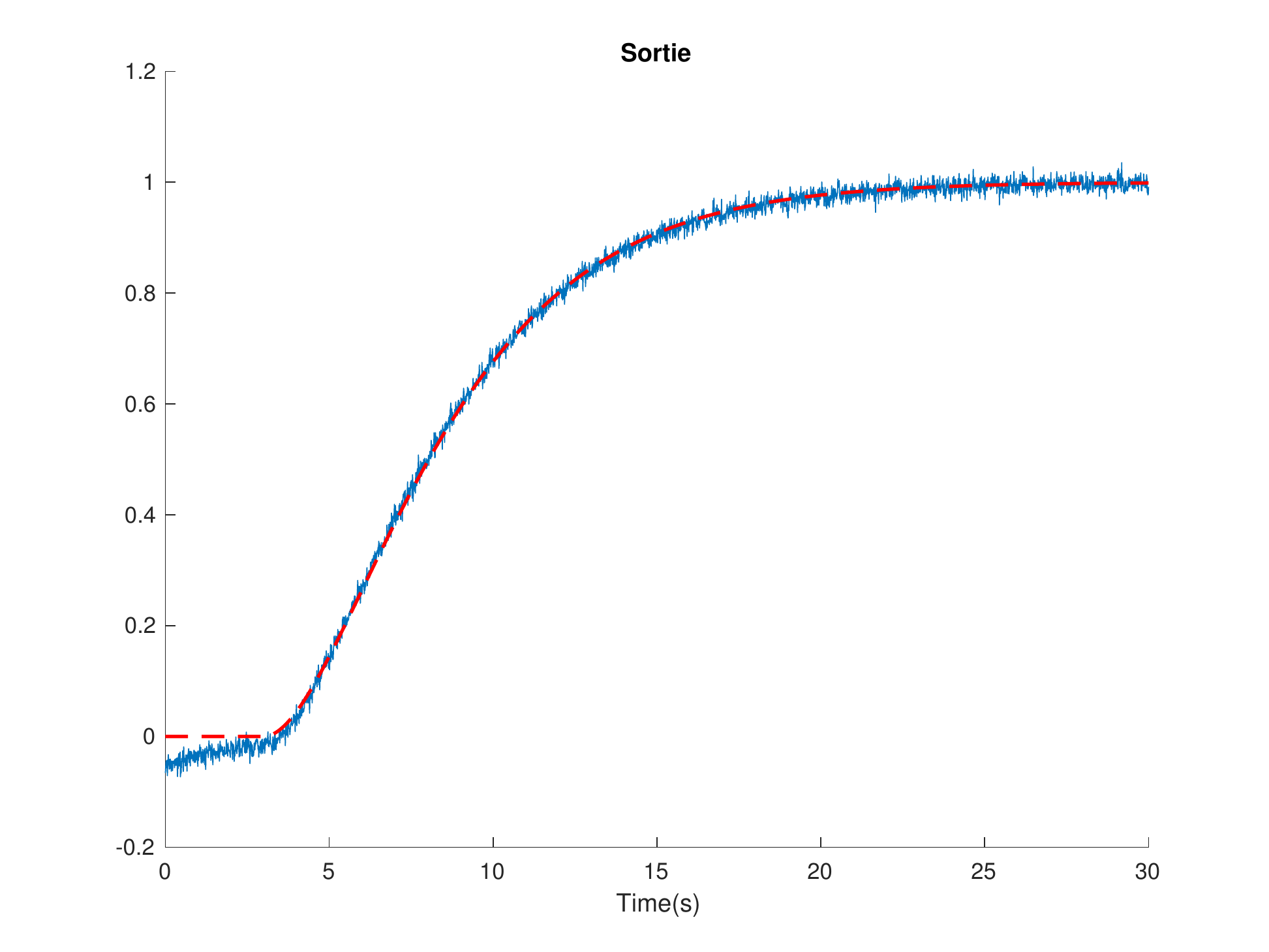}}}%
\caption{iPD}%
\label{ipd}
\end{center}
\end{figure*}

Avec un iPD \eqref{iPD}, $\nu = 2$ en \eqref{1}, on remplace \eqref{Fana} par
\begin{equation*}\label{2}
F=(1-\alpha)\ddot{y} + \alpha \dot{y}
\end{equation*}
Les simulations numériques de la figure {\ref{ipd}}, déduites des calculs de \cite{ijc13}, sont excellentes. On choisit $\alpha =0.5$ et les gains $K_P$ and $K_D$ tels que $(s + 0.5)^2$ est le polynôme caractéristique de la dynamique d'erreur. On introduit un bruit additif de sortie, blanc, centré et gaussien, d'écart type $0.01$. La condition initiale est $y(0)=-0.05$.

\section{Pourquoi l'implantation de l'iP échoue-t-elle?}\label{IP}
\begin{figure*}
\begin{center}
\subfigure[Pour tout $T$]{
\resizebox*{7.80cm}{!}{\includegraphics{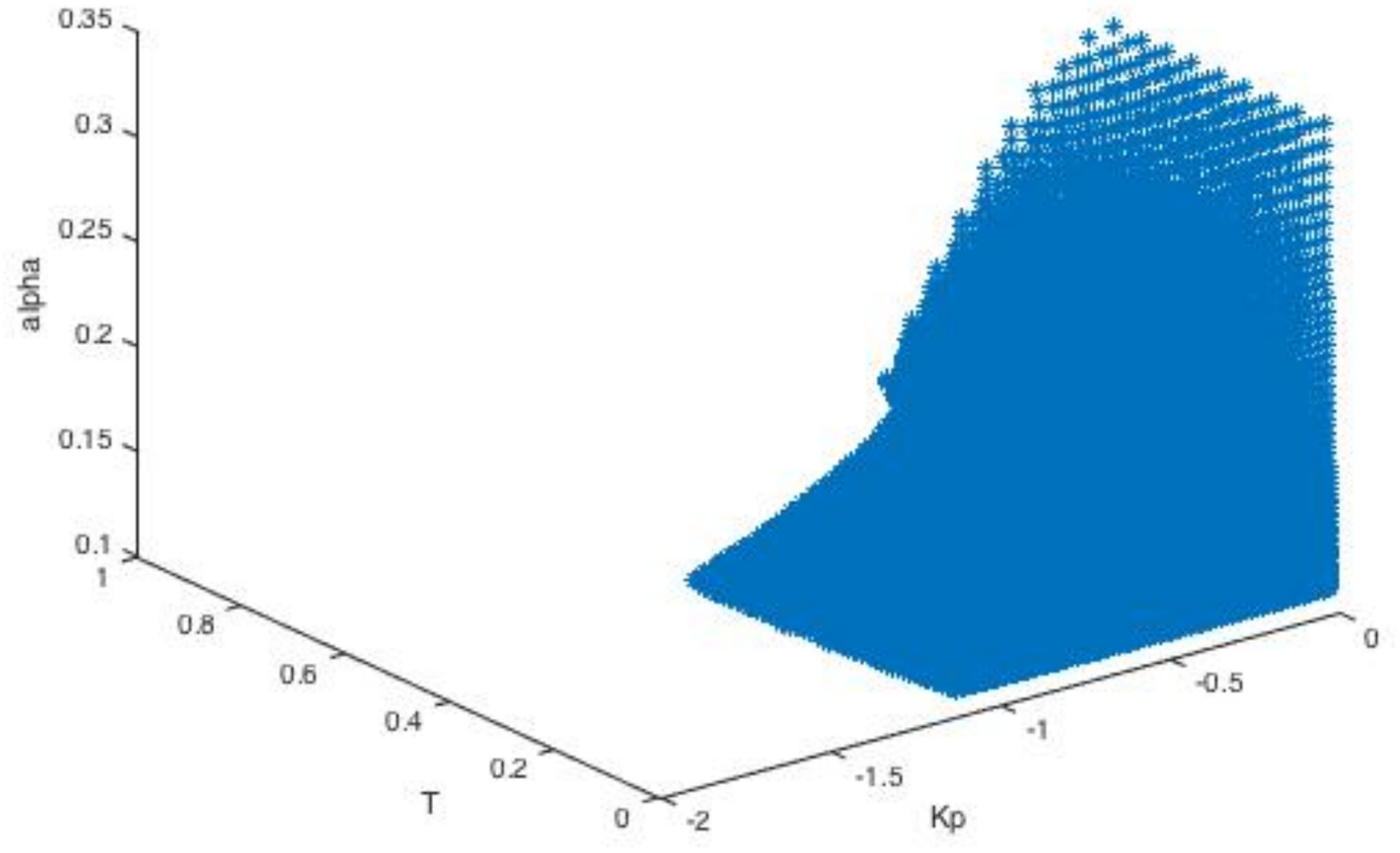}}}%
\subfigure[$T=0.1$s]{
\resizebox*{7.80cm}{!}{\includegraphics{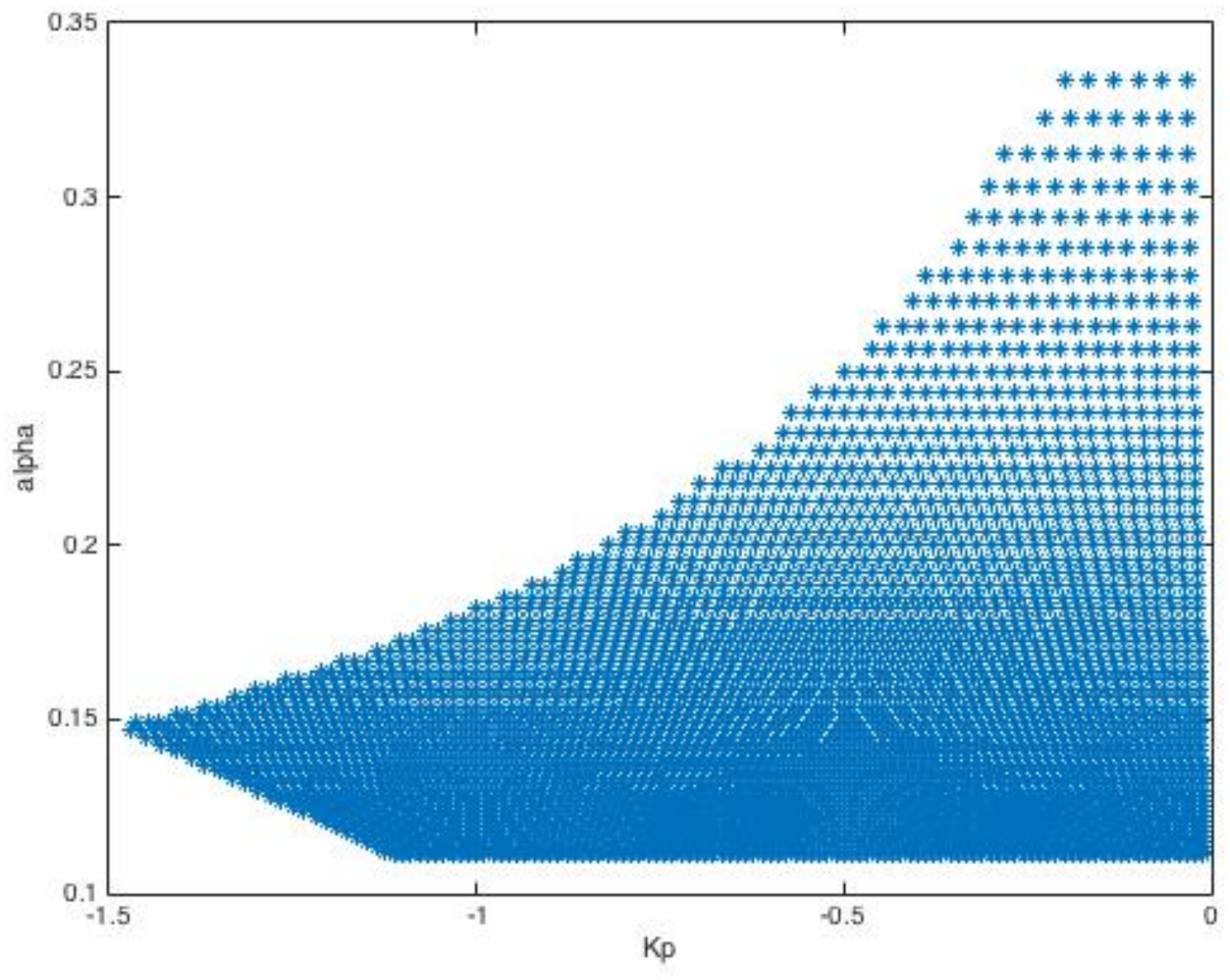}}}%
\caption{ Domaine de stabilité pour $(K_P, \alpha, T)$ }%
\label{Sip}
\end{center}
\end{figure*}

Tentons de comprendre l'échec de l'implantation de l'iP. 
Il vient pour \eqref{track}:
\begin{equation}\label{track1}
\dot{e} + K_P e = F_{\rm estim} - F
\end{equation}
où $F$ provient de \eqref{Fana}. Il est loisible d'écrire $F_{\rm estim}$ dans le domaine opérationnel (voir, par exemple, \cite{yosida}):
\begin{equation}\label{Fanad}
{\frak F}_{\rm estim} =-\alpha \frac{s^2}{(Ts+1)^2}{\frak y} +(1+\alpha)\frac{s}{(Ts+1)}{\frak y}
\end{equation}
\begin{itemize}
\item ${\frak F}_{\rm estim}$ et ${\frak{y}}$ sont les analogues opérationnels\footnote{La terminologie \og{transformées de Laplace}\fg \ est beaucoup plus usuelle, comme chacun le sait.} de $F_{\rm estim}$ et $y$,
\item $\frac{s}{(Ts+1)}$ et $\frac{s^2}{(Ts+1)^2}$ représentent des filtres dérivateurs d'ordres $1$ et $2$, où $T > 0$ est la constante de temps (voir, par exemple, \cite{filter}).
\end{itemize}
Il est loisible de poser $y^\ast \equiv 0$ pour étudier la stabilité. Alors, $e = - y$. Gr\^{a}ce à \eqref{x}-\eqref{Fanad}, on obtient le polynôme caractéristique
%
%
$$
T^2s^4+s^3(2T-T^2) +s^2\left(-2T+T(1+\frac{1}{\alpha})-T^2\frac{K_P}{\alpha} \right)
+s \left(\frac{1}{\alpha}-2T\frac{K_P}{\alpha}\right) - \frac{K_P}{\alpha}
$$
On utilise le critère de stabilité de Routh-Hurwitz (voir, par exemple, \cite{gant}, et \\ \cite{franklin}, \cite{lunze}, \cite{rotella}). La figure {\ref{Sip}} montre, avec une discrétisation convenable, l'étroitesse de l'ensemble $\{\alpha, K_P\}$ des paramètres stabilisants, y compris en tenant compte de la constante de temps $T$.  La difficulté de trouver un iP satisfaisant pour \eqref{x} se trouve confirmée. Ajoutons que la valeur \og{évidente}\fg \ $\alpha = - 1$ ne convient jamais\footnote{La valeur $\alpha = - 1$ est \og{évidente}\fg \ car, alors, $F = \ddot{y}$ en \eqref{Fana}. On retrouve \eqref{x}.}.

%


\section{Comparaison entre iPD et PID}\label{PD}
\begin{figure*}
\begin{center}
\subfigure[Commande]{
\resizebox*{7.80cm}{!}{\includegraphics{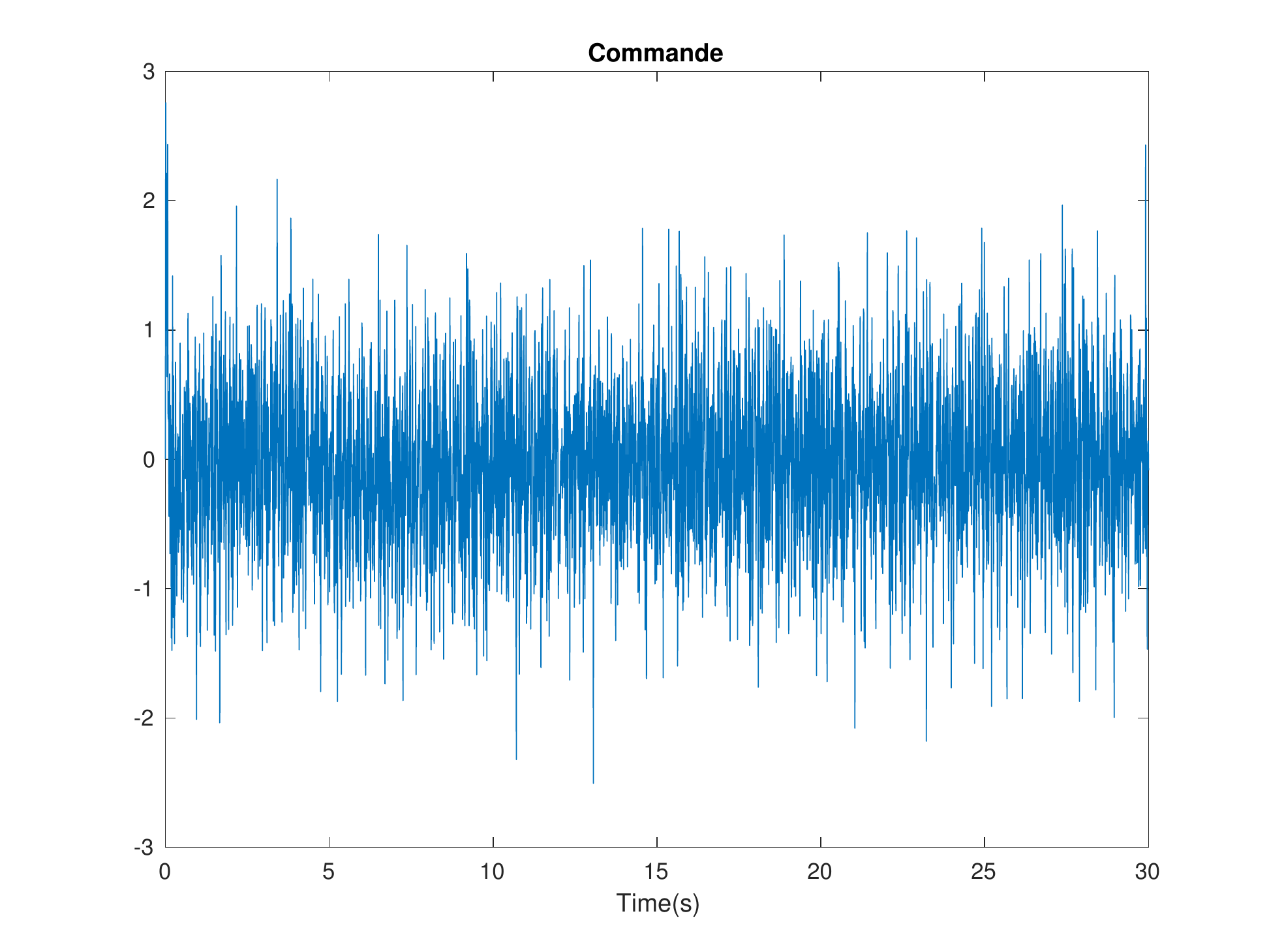}}}%
\subfigure[Sortie, trajectoire de référence (- -)]{
\resizebox*{7.80cm}{!}{\includegraphics{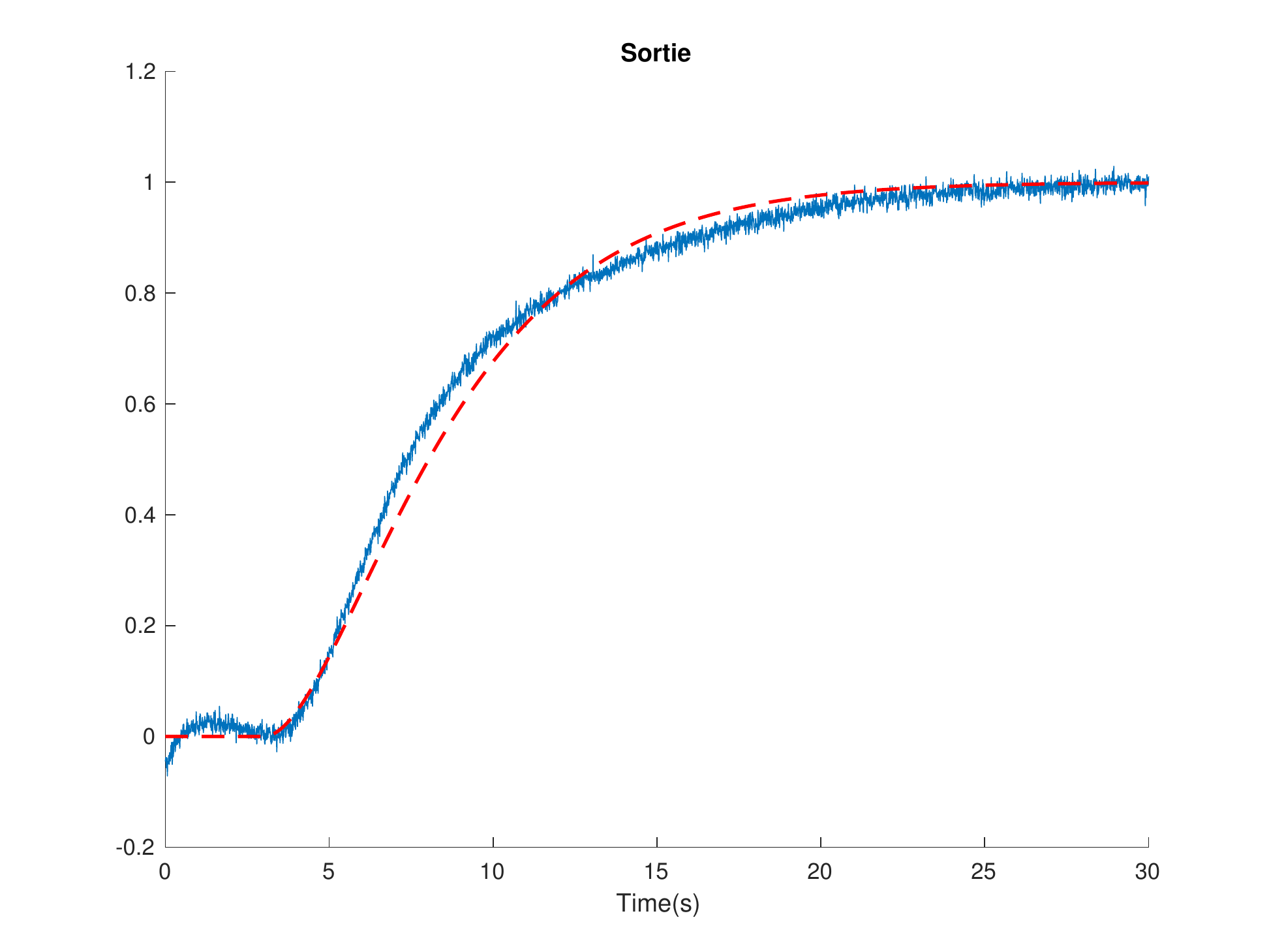}}}%
\caption{PID}%
\label{PID}
\end{center}
\end{figure*}

\begin{figure*}
\begin{center}
\subfigure[Commande]{
\resizebox*{7.80cm}{!}{\includegraphics{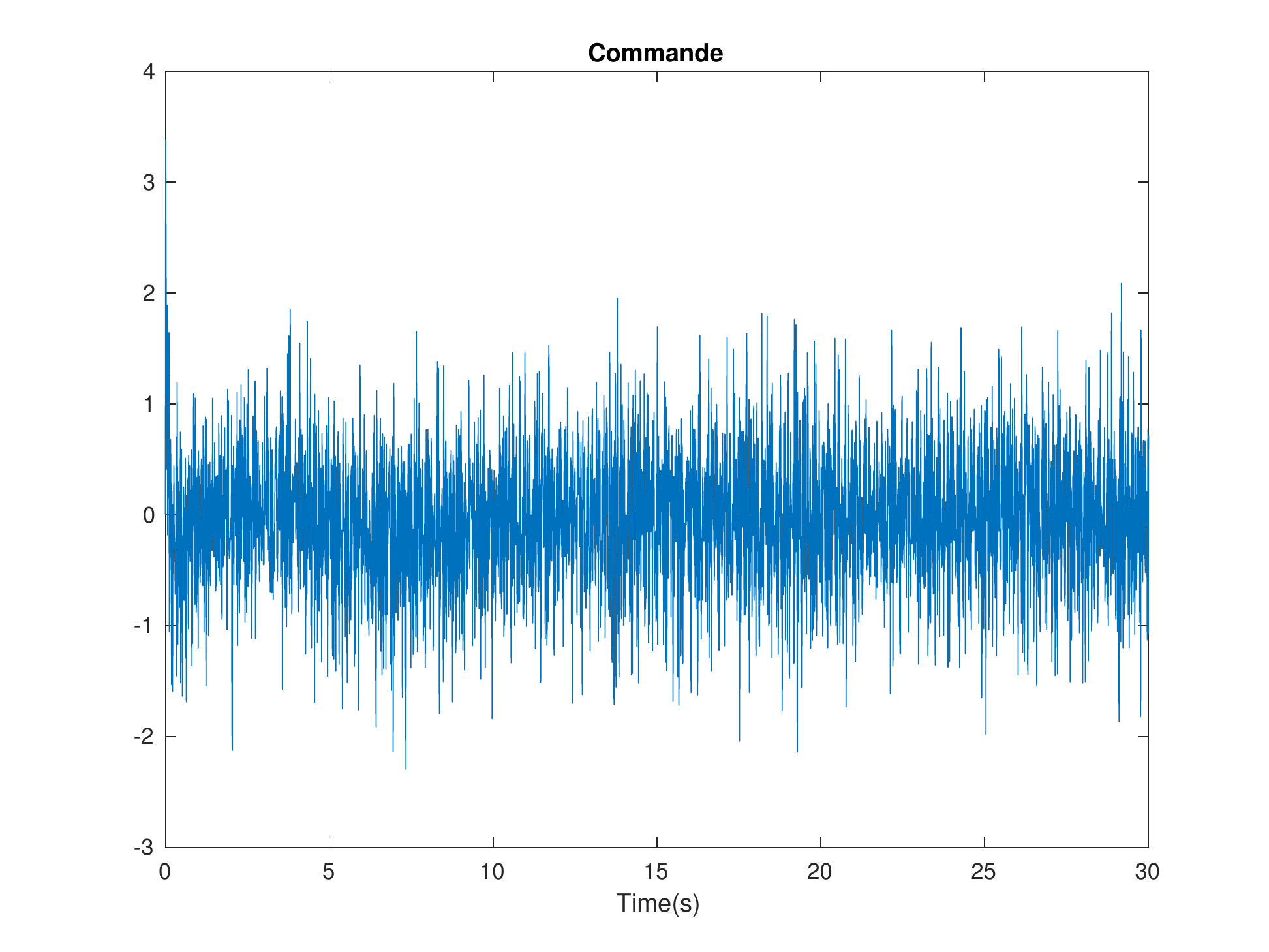}}}%
\subfigure[Sortie, trajectoire de référence (- -)]{
\resizebox*{7.80cm}{!}{\includegraphics{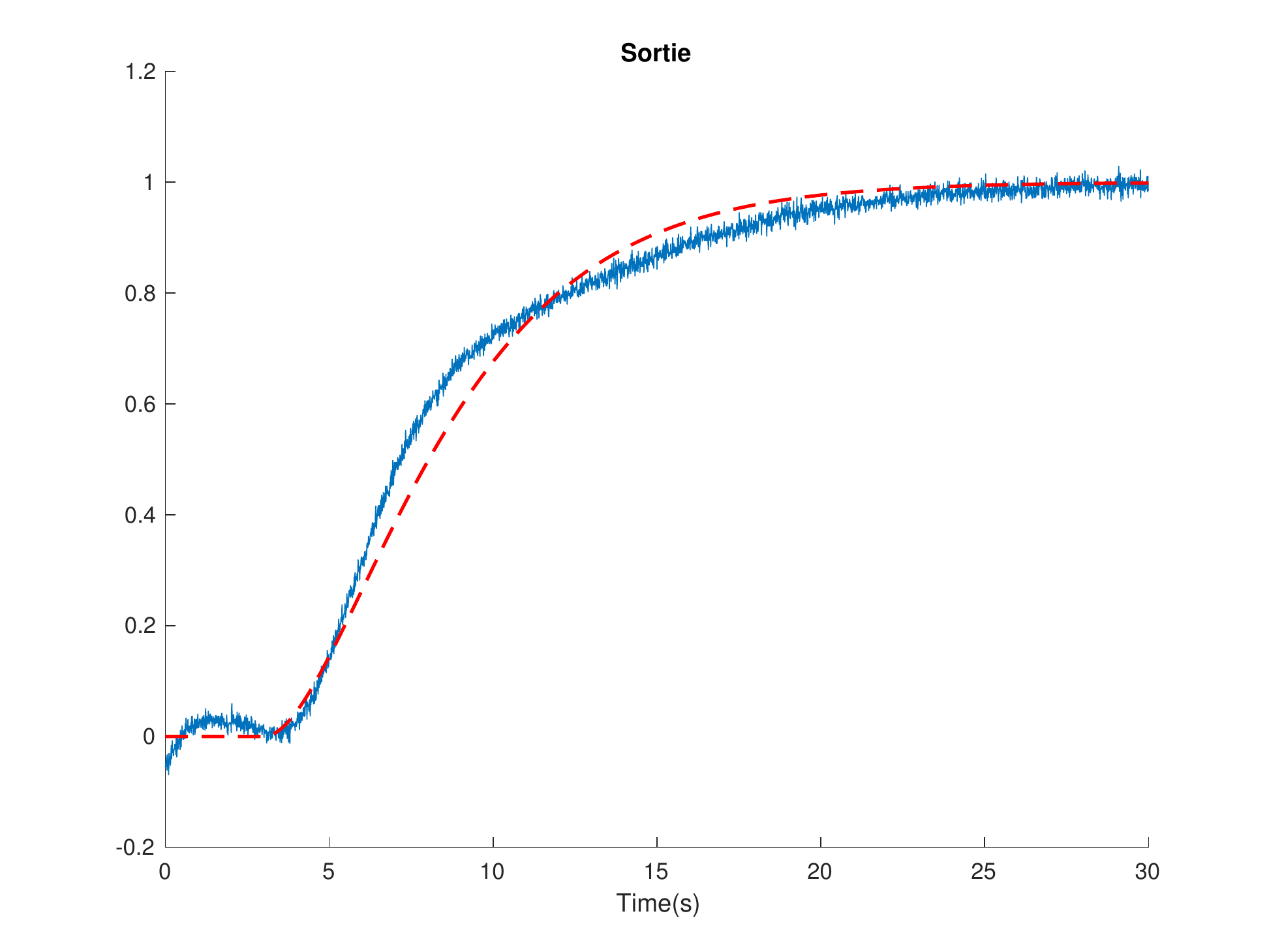}}}%
\caption{PID avec $\delta=0.8$}%
\label{PID8}
\end{center}
\end{figure*}

\begin{figure*}
\begin{center}
\subfigure[Commande]{
\resizebox*{7.80cm}{!}{\includegraphics{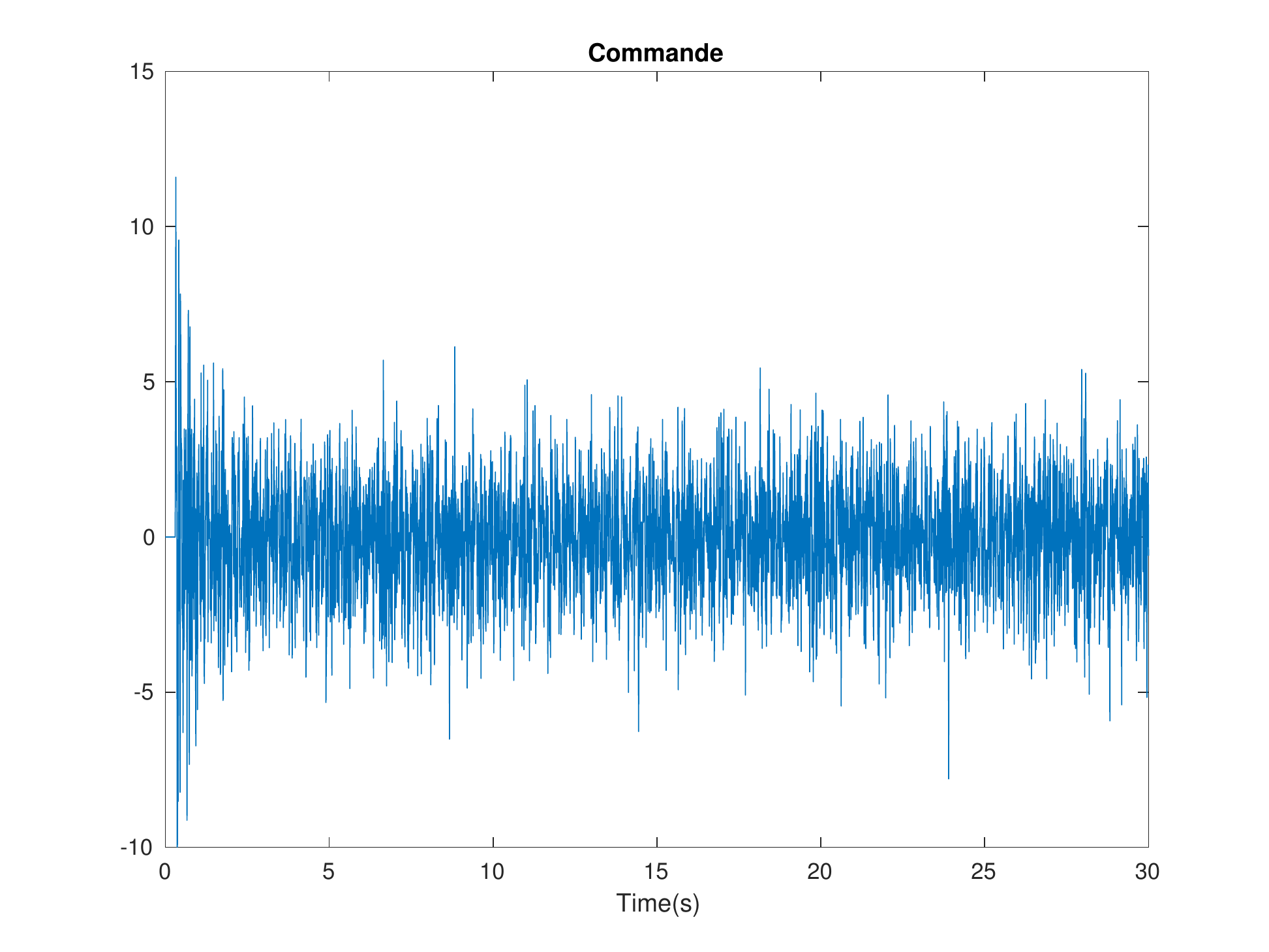}}}%
\subfigure[Sortie, trajectoire de référence (- -)]{
\resizebox*{7.80cm}{!}{\includegraphics{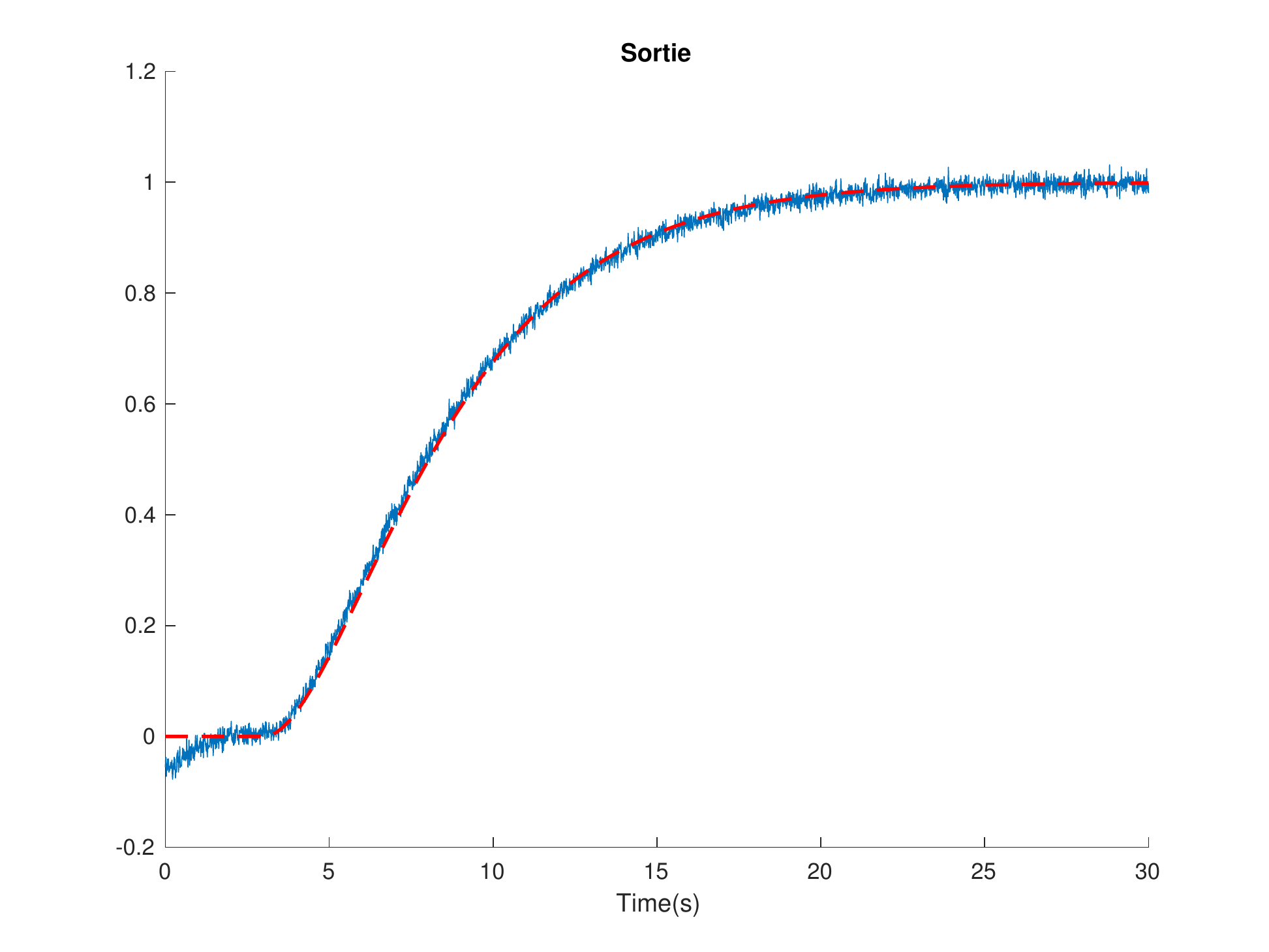}}}%
\caption{iPD avec $\delta=0.8$}%
\label{iPD8}
\end{center}
\end{figure*}

\begin{figure*}
\begin{center}
\subfigure[Commande]{
\resizebox*{7.80cm}{!}{\includegraphics{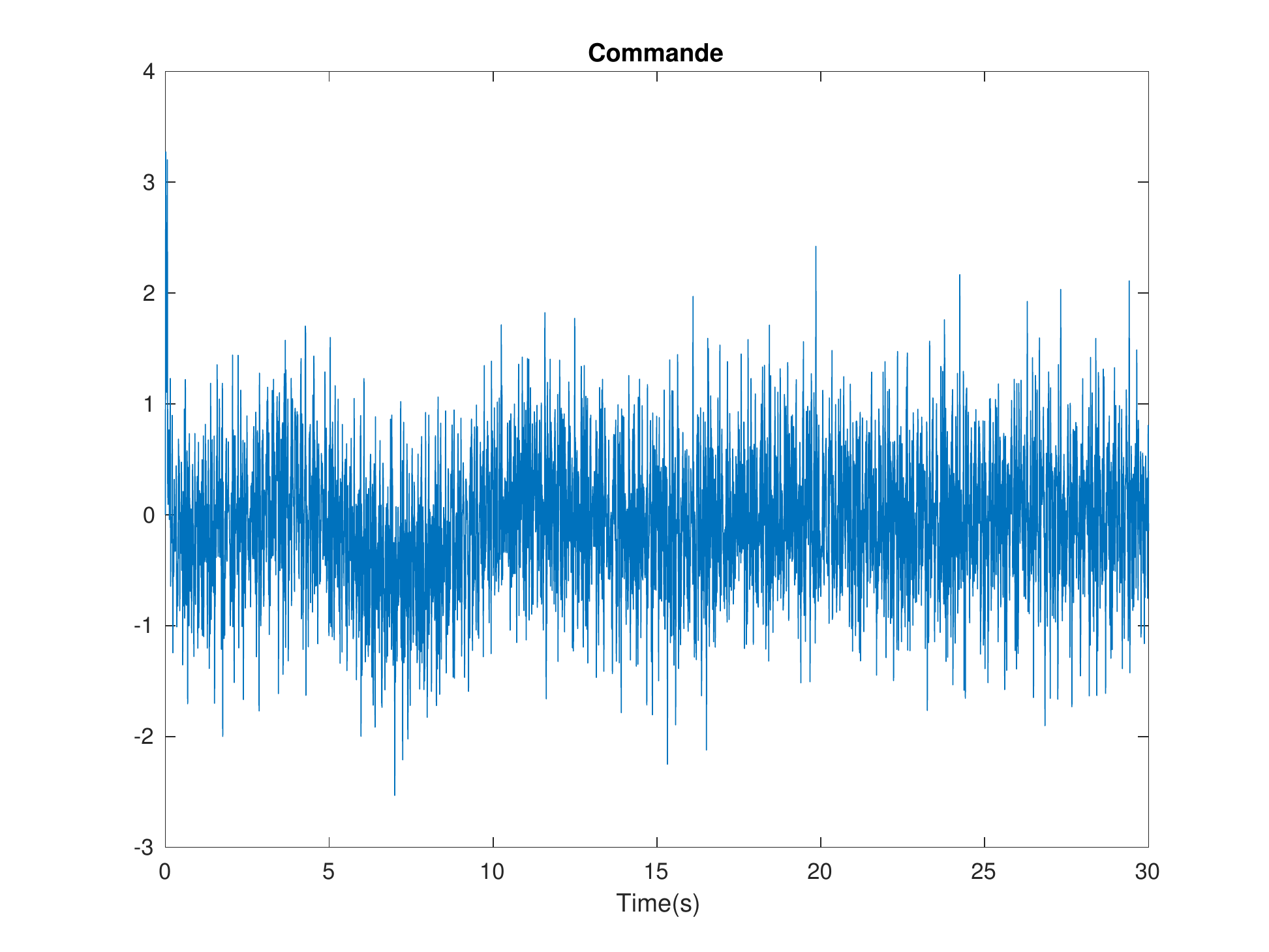}}}%
\subfigure[Sortie, trajectoire de référence (- -)]{
\resizebox*{7.80cm}{!}{\includegraphics{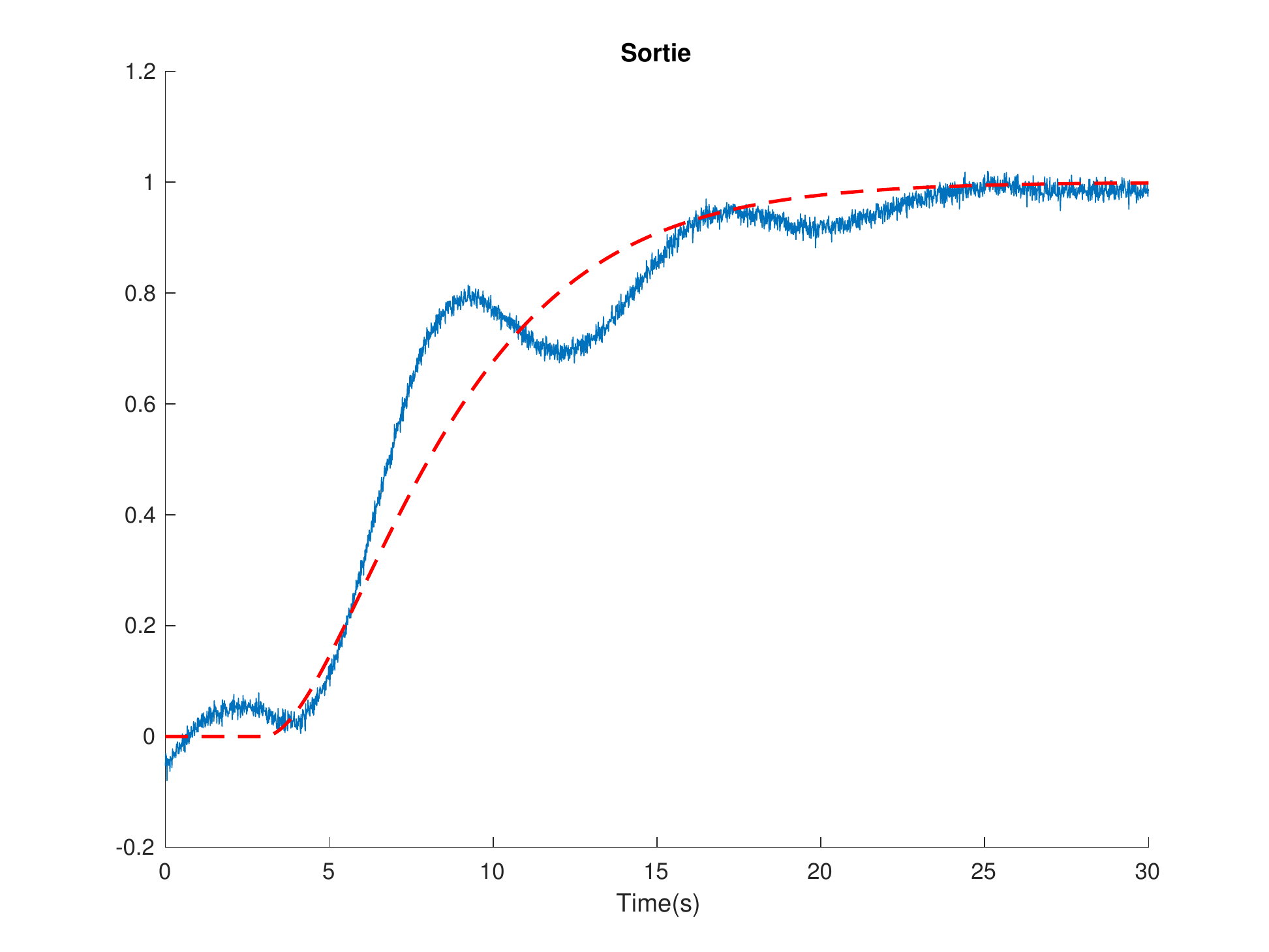}}}%
\caption{PID avec $\delta=0.5$}%
\label{PID5}
\end{center}
\end{figure*}

\begin{figure*}
\begin{center}
\subfigure[Commande]{
\resizebox*{7.80cm}{!}{\includegraphics{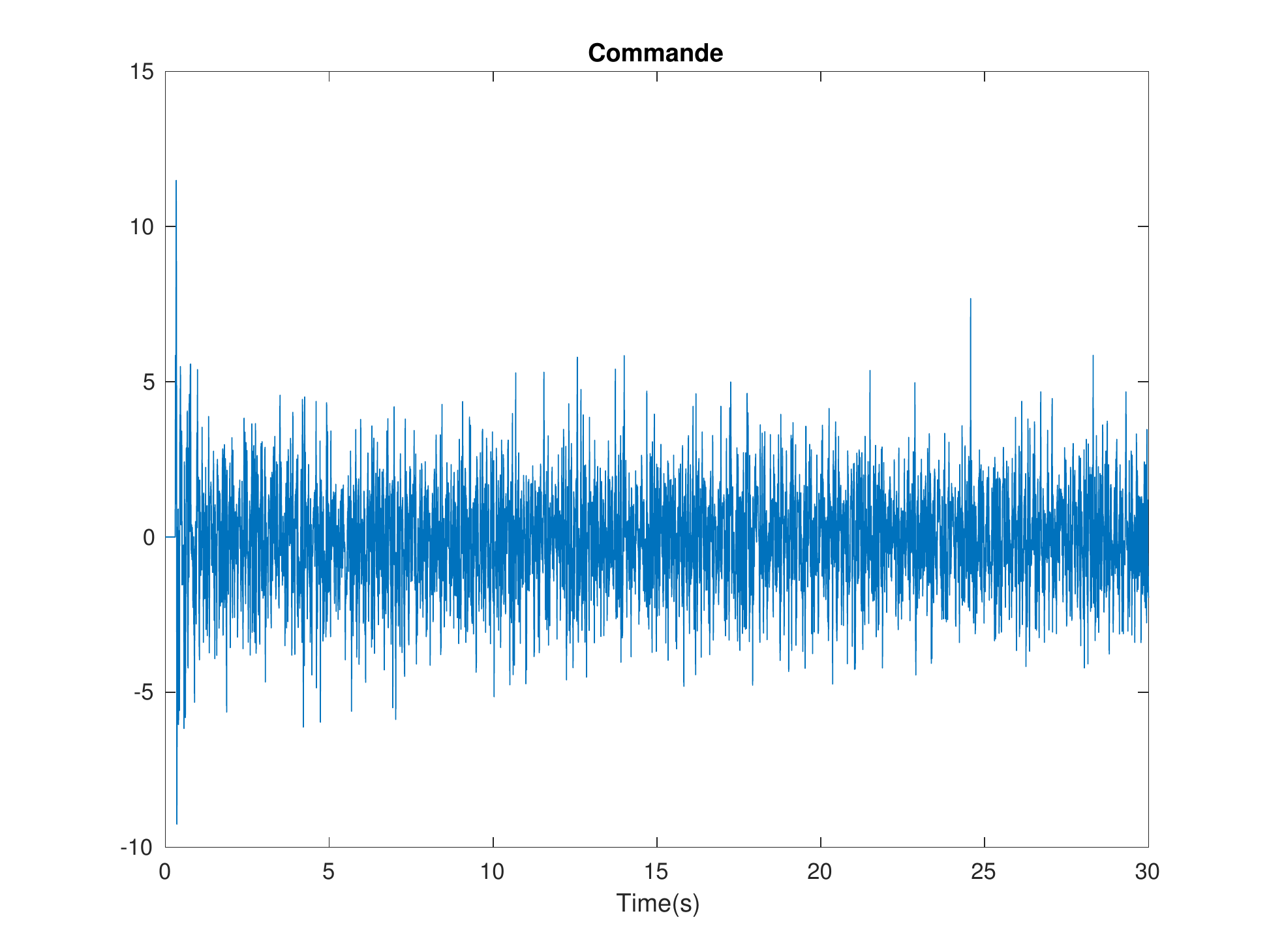}}}%
\subfigure[Sortie, trajectoire de référence (- -)]{
\resizebox*{7.80cm}{!}{\includegraphics{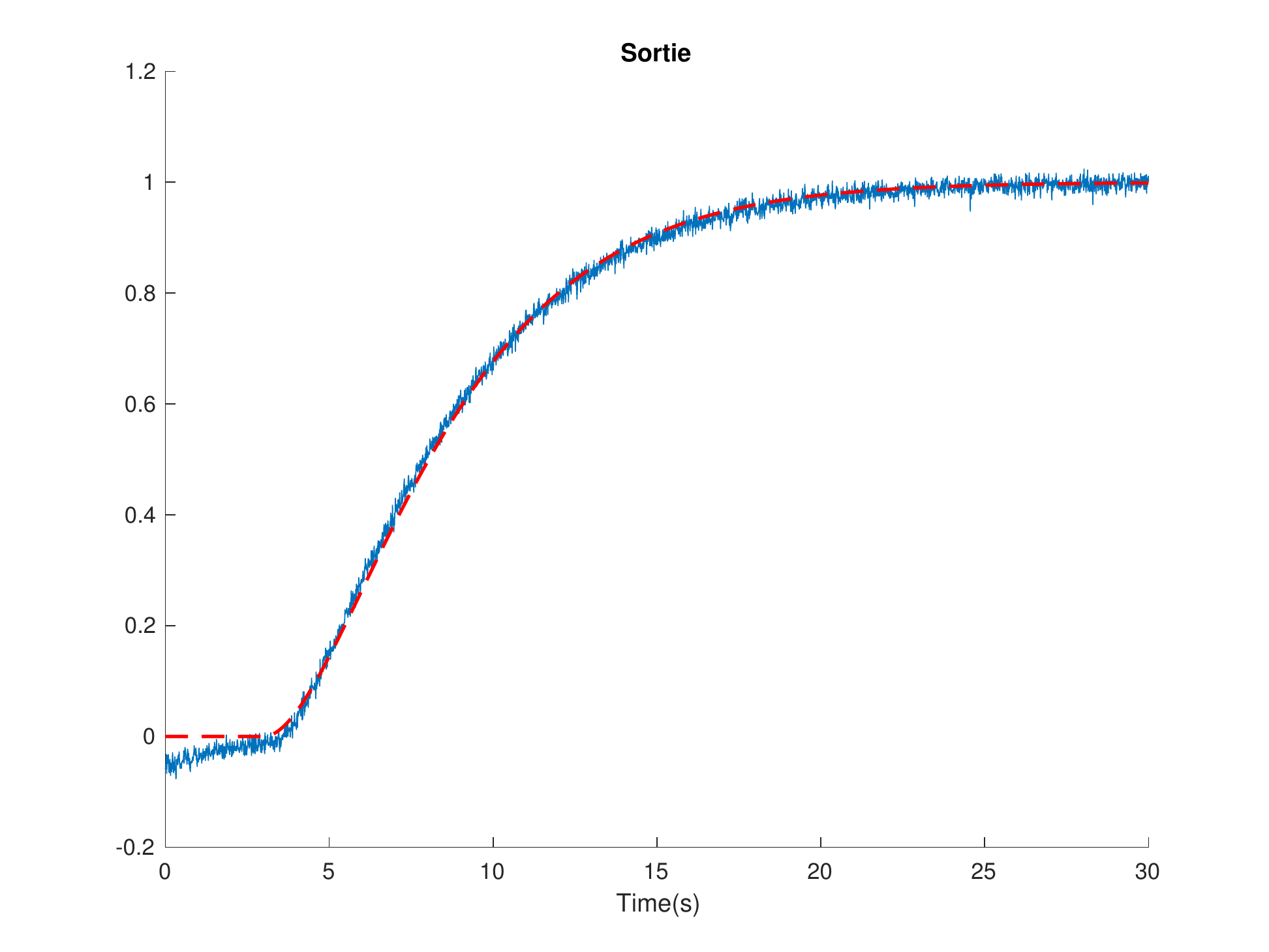}}}%
\caption{iPD avec $\delta=0.5$}%
\label{iPD5}
\end{center}
\end{figure*}

En \cite{pId}, \cite{ijc13} est démontrée une certaine équivalence entre iPD \eqref{iPD} et PID usuels:
\begin{equation*}\label{PIDc}
u = k_P e + k_I \int e + k_D \dot{e}   \quad (k_P, k_I, k_D \in \mathbb{R})
\end{equation*} 
On détermine $k_P$, $k_I$, $k_D$ de sorte que $(s + 0.66)^3$ soit le polynôme caractéristique de la dynamique d'erreur. On assure ainsi un temps de réponse égal à celui du paragraphe {\ref{IPD}}, à $\pm 5\%$ près. Les résultats de la figure {\ref{PID}}  sont satisfaisants\footnote{Un réglage plus soigneux du PID les améliorerait peut-être.} quoiqu'inférieurs à ceux de la figure {\ref{ipd}} pour l'iPD\footnote{Même bruit qu'au paragraphe {\ref{IPD}}}. 
Afin de tester la robustesse, modifions \eqref{x}:
\begin{equation*}\label{xd}
{\ddot{y}-\dot y=\delta u }
\end{equation*}
où $\delta$, $0 \leq \delta \leq 1$, correspond à une perte de puissance de l'actionneur\footnote{Comparer avec la \og{réparation}\fg, ou \emph{fault accommodation}, en \cite{ijc13}. Voir, aussi, \cite{toulon}.}. Avec $\delta = 0.8$, figures {\ref{PID8}} and {\ref{iPD8}}  révèlent un meilleur comportement de l'iPD. Cette supériorité s'accentue si $\delta = 0.5$: voir figures {\ref{PID5}} et {\ref{iPD5}}.


\section{Conclusion}\label{concl}
Diverses questions découlent de cette étude:
\begin{enumerate}
\item L'iP et l'iPD sont les deux seuls correcteurs intelligents qui devraient importer. Les autres, comme iPI et iPID, ne joueront, sans doute, qu'un rôle marginal.
\item L'iPD s'impose ici grâce à l'examen d'une équation donnée. Exhiber d'autres équations de systèmes jouissant de cette propriété est un but intellectuel légitime. Mais que faire sans modèle? Procéder par essais et erreurs ? Tenir compte des propriétés qualitatives de la machine ? Un mélange des deux? Voilà encore une interrogation épistémologique nouvelle, due au surgissement de la commande sans modèle\footnote{\`A ce sujet, voir aussi la conclusion de \cite{ijc13}.}.
\item La preuve repose sur le choix des filtres dérivateurs. Une analyse indépendante d'une telle approche reste à découvrir. Elle permettrait une meilleure compréhension du phénomène.
\end{enumerate}

\end{document}